\documentclass[11pt,twoside]{article}
\usepackage{ntsha_win}
\usepackage[OT1,T2A]{fontenc}
\usepackage[cp1251]{inputenc}
\usepackage[english,ukrainian]{babel}
\renewenvironment{abstract}{\begin{quotation}\small
\mdseries}{\end{quotation}
}
\addto\captionsukrainian{%
}

\usepackage[tbtags]{amsmath}
\usepackage{cite,amssymb,amsthm,dsfont
}
\usepackage[mathscr]{eucal}
\title[Релятивіська дзиґа]{ВАРІЯЦІЙНІ РІВНЯННЯ ТРЕТЬОГО ПОРЯДКУ ДЛЯ РЕЛЯТИВІСЬКОЇ ДЗИҐИ}
\author[Р.~Мацюк]
{Роман~МАЦЮК}
\date{Інститут прикладних проблем механіки і математики\\ НАН України,\\
 вул.~Наукова,~3\textsuperscriptб, Львів 79000
 \footnotetext{PACS~2006~numbers 11.15.Kc, 02.40.Ky, 45.20.Jj, 45.50.-j}
\footnotetext{Стаття подається в авторській редакції}}
\newtheorem*{lema}{Лема}
\newtheorem{prop}{Річ}
\theoremstyle{remark}
\newtheorem*{zauv}{Зауваження}
\begin{document}
\setcounter{page}{77}  %   заповнюється редакцією
{\renewcommand{\baselinestretch}{1.2}
 \maketitle
}
 {\hfill \small Редакція отримала статтю 20~листопада~2012~р.}
%   заповнюється редакцією
\thispagestyle{myheadings}
\markboth{~\hrulefill~Фізичний збірник НТШ т.",9 2014~p.}
     {Фізичний збірник НТШ т.",9 2014 p.~\hrulefill~}
\begin{abstract}
Подаємо виведення рівняння руху третього порядку для вільної релятивіської дзиґи із засади варіяційности в поєднанні з вимогою його самозмінности щодо перетворень Лоренца.
\end{abstract}
{
\newcommand{\uap}{{\kern-.1em'}}
\newcommand{\uapi}{{\kern-.1em'\kern.07em}}
\newcommand{\SSS}{\scriptscriptstyle}
\newcommand{\x}{{\mathsf x}}
\newcommand{\sfs}{{s}}
\newcommand{\sfe}{\boldsymbol{\mathsf e}}
\newcommand{\sv}{{\mathsf v}}
\newcommand{\sE}{{\mathsf E}}
\let\E=\sE
\newcommand{\se}{\alpha}
\newcommand{\BBN}{\mathds N}
\newcommand{\bmat}[1]{{\boldsymbol#1}}
\newcommand{\sss}[1]{{\scriptscriptstyle#1}}
\newcommand\so{{\scriptscriptstyle 0}}
\newcommand\R{{\mathds R}}
\newcommand\sA{\mathsf A}
\newcommand\sB{\mathsf B}
\newcommand\sC{\mathsf c}
\newcommand\sk{\mathsf k}
\newcommand{\A}{\boldsymbol{\mathsf A}}
\newcommand{\B}{\boldsymbol{\mathsf B}}
\newcommand{\G}{\boldsymbol{\mathsf G}}
\newcommand{\CC}{\boldsymbol{\mathsf  c}}
\newcommand{\V}{\boldsymbol{\mathsf  v}}
\newcommand{\X}{\boldsymbol{\mathsf  x}}
\newcommand{\K}{\boldsymbol{\mathsf  k}}
\newcommand{\n}{\boldsymbol{\mathsf  n}}
\newcommand{\z}{\boldsymbol{\mathsf  z}}
\newcommand{\W}{\boldsymbol{\Omega}}
\renewcommand{\a}{\boldsymbol{\mathfrak a}}
\newcommand{\fa}{\boldsymbol{\mathfrak a}}
\newcommand{\bP}{\boldsymbol{\pi}}
\newcommand{\w}{\boldsymbol{\omega}}
\newcommand{\Zero}{\boldsymbol{\mathsf  0}}
\newcommand{\bcdot}{\boldsymbol{\cdot}}
\newcommand{\bpartial}{\boldsymbol{\partial}}
\newcommand{\bkey}[1]{\boldsymbol{#1}}
\newcommand{\bdot}{\boldsymbol{\,.\,}}
\newcommand{\rA}{\mathscr A}
\newcommand{\rB}{\mathscr B}
\newcommand{\rC}{{\scriptstyle\mathscr C}}
\newcommand{\bA}{\boldsymbol{\mathscr A}}
\newcommand{\bB}{\boldsymbol{\mathscr B}}
\newcommand{\bC}{\boldsymbol{{\scriptstyle\mathscr C}}}
\newcommand{\bU}{\boldsymbol{u}}
\newcommand{\bX}{\boldsymbol{x}}
\newcommand{\bs}{\boldsymbol{s}}
\newcommand{\bbs}{{\scriptstyle\mathbb S}}
\newcommand{\bbss}{{\scriptscriptstyle\mathbb S}}
\newcommand{\iu}{\mathit u}
\newcommand{\ix}{\mathit x}
\newcommand{\ba}{\boldsymbol{\mathsf a}}
\newcommand{\bb}{\boldsymbol{\mathsf b}}
\newcommand{\ff}{\boldsymbol{\mathfrak f}}
\providecommand{\norm}[1]{\lVert#1\rVert}
\providecommand{\msp}{{\kern.03em}}
\providecommand{\hstrut}{{\vrule height 7pt width 0pt depth 0pt}}
\providecommand{\dstrut}{{\vrule height 0pt width 0pt depth 4pt}}
\section{Вступ.}
Ця праця покликана доповнити попередні дослідження про варіяційність рівнянь руху, що містять третю похідну від координати частки, переведені у статтях~\cite{Matsyuk:CMP1998,Matsyuk:NTSh2006,matsyuk:MatMet1984}.
Наприкінці попередньої розвідки~\cite{Matsyuk:NTSh2006}
встановлено, що рівняння
\begin{equation}\label{matsyuk:26}
{\frac{{ *\, {\boldsymbol{\ddot {u}}} \wedge {\boldsymbol{{u}}} \wedge {\boldsymbol{s}}}}{{{\left\| {{\boldsymbol{s}} \wedge {\boldsymbol{{u}}}}
\right\|}^{3}}}} - 3\,{\frac{{\left( {{\boldsymbol{s}} \wedge {\boldsymbol{u}}} \right) \cdot \left( {{\boldsymbol{s}} \wedge {\boldsymbol{\dot {u}}}}
\right)}}{{{\left\| {{\boldsymbol{s}} \wedge {\boldsymbol{{u}}}}
\right\|}^{5}}}}\; * {\boldsymbol{\dot {u}}} \wedge {\boldsymbol{{u}}} \wedge {\boldsymbol{s}} + {\frac{{\mu} }{{{\left\| {{\boldsymbol{s}}}
\right\|}^{3}{\left\| {{\boldsymbol{{u}}}} \right\|}^{3}}}}\left( {{\boldsymbol{u}}^{2}{\boldsymbol{\dot {u}}} - \left( {{\boldsymbol{{u}}}
\cdot {\boldsymbol{\dot {u}}}} \right)\,{\boldsymbol{{u}}}} \right) = 0
\end{equation}
при постійних $\mu$  та ${\boldsymbol{s}}$ описує рух вільної
 дзиґи з масою
\begin{equation}\label{matsyuk:7}
m = \mu \left( {1 - {\frac{{\left( {{\boldsymbol{s}} \cdot {\boldsymbol{{u}}}}
\right)^{2}}}{{{\boldsymbol{s}}^{2}{\boldsymbol{{u}}}^{2}}}}}
\right)^{3/2},
\end{equation}
де значком ${\boldsymbol{s}}$ позначений деякий чотири-вектор
неправдивого \glqq спіну\grqq{} (\emph{крутня}) цієї дзиґи.
Величина
 \begin{equation}\label{matsyuk:6}
{\frac{{{\boldsymbol{s}} \cdot {\boldsymbol{u}}}}{{{\left\| {{\boldsymbol{u}}}
\right\|}}}}
\end{equation}
 є першим інтеґралом рівняння~(\ref{matsyuk:26}).
У даній розвідці розглядатимемо самозмінні щодо групи рухів варіяційні рівняння у чотиривимірному просторі спеціяльної теорії відносности. Зокрема, покажемо, що рівняння~(\ref{matsyuk:26}) можна отримати з варіяційної засади.
Як технічну передумову, розширимо правдоміру варіяційности, описану в розвідці~\cite{Matsyuk:CMP1998} для рівнянь у часовому в\'ідмірі, на випадок теорії в одноманітних ("`однорідних"') співрядних. Докладно розглянемо пов\uapязання і відповідність між мовою та технічним апаратом дослідження оберненого варіяційного завдання в часовій в\'ідмірності та у формалізмі в\'ідмірної байдужости.
\section{Варіяційність і в\'ідмірна байдужість}
Рівняння~(\ref{matsyuk:26}) є параметрично-інваріянтним (ще кажуть \emph{в\'ідмірно\cdash--~байдужим}, або \emph{безвідмірним}) щодо зміни параметра (\emph{в\'ідміру}) уздовж його інтеґральних стежок. Для запису подібних безвідмірних сутностей, замість так званих \emph{одноманітних співрядних}, \cdash--- швидкостей $u^\alpha$, $\dot u^\alpha$, $\ddot u^\alpha$, \dots,
$u_{\dstrut(r)}^\alpha$, \cdash--- можна використовувати \emph{співрядні торкання}, \cdash---  швидкості $\sv\msp^i$, $\sv\msp'\msp^i$, $\sv\msp''\msp^{i}$,\dots, $\sv_{(r)}^{i}$, віднесені до змінної $t=x^{\SSS0}$. В третьому порядку перерахунок між змінними відбувається згідно зі взором:
\begin{eqnarray}
&&\sv^i    = \frac1{\dot t^{\strut}}\, u^i  \nonumber
\\ && {\sv'}\msp^i  = \frac1{(\dot t)^{\strut 3}}  \left(\dot t \dot u^i - \ddot t u^i\right) \label{matsy:calp}
\\ &&{\sv''}\msp^i  =\frac1{(\dot t^{\strut})^{5}}  \left\{(\dot t)^2 \ddot u^i - 3\, \dot t\; \ddot t \,\dot
u^i + \left[3(\ddot t)^2 -\dot t \;\dddot t\right] u^i\right\}.   \nonumber
\end{eqnarray}
Позначимо буквою $p^r$ перехід від змінних  $u^\alpha$, $\dot u^\alpha$, $\ddot u^\alpha$, \dots,
$u_{(r)}^\alpha$ до змінних $\sv^i$, $\sv'^i$, $\sv''^i$,\dots, $\sv_{(r)}^i$.

Нехай у змінних $\sv\msp^i$, $\sv'\msp^i$, $\sv''\msp^i$,\dots, $\sv_{(k)}^i$ задана деяка ляґранжева густина
\begin{equation}
\Lambda=L\left(t; \x^i, \sv^i, {\sv'}^i, {\sv''}^i, \dots, \sv_{(k)}^i \right)dt\,,
\label{matsy:L}
\end{equation}
якій відповідає варіяційне рівняння
\begin{equation}
\sE_i\left(t; \x^i, \sv^i, {\sv'}^i, {\sv''}^i, \dots, \sv_{(s)}^i \right) = 0\,.
\label{matsy:fe}
\end{equation}
Нехай, далі, у змінних $\sv\msp^i$, $\sv'\msp^i$, $\sv''\msp^i$,\dots, $\sv_{(k)}^i$ поставлено варіяційне завдання з функцією Ляґранжа
\begin{equation}
{\mathcal L}=\dot t \;\big( L\circ p^k\big)\,.
\label{matsy:L0}
\end{equation}

Попереднім дослідженням~\cite{Matsyuk:IPPMM}
 встановлена така правда:
\begin{prop}\label{matsy:homogen}
Якщо варіяці\-йне рівняння (\ref{matsy:fe}) відповідає ляґранжевій густині~(\ref{matsy:L}), то варіяційне рівняння
\begin{equation}\label{matsy:CalE}
    \{{\mathcal E}_\alpha\} \overset{\mathrm{def}}= \left(
                                \begin{array}{r}
                                  -\,u^i\,\sE_i\circ p^s \\
                                   \dot t \;\sE_i\circ p^s\\
                                \end{array}
                              \right)
=0
\end{equation}
відповідає функції Ляґранжа (\ref{matsy:L0}).
\end{prop}

Рівняння (\ref{matsy:CalE}) порядку~$s+1$ описує в
\glqq однорідному\grqq{} вигляді ті ж безвідмірні інтеґральні стежки, керовані варіяційним  завданням~(\ref{matsy:L0}), що й рівняння
(\ref{matsy:fe}).
Функція Ляґранжа ${\mathcal L}$
вочевидь задовольняє так звані умови Цермело, які є критерієм параметричної байдужости відповідного варіяційного завдання. В другому порядку ці умови записуються ось як:
\begin{eqnarray*}
   &   & u^\beta\frac{\partial}{\partial u^\beta}{\mathcal L}+2\dot
u^\beta\frac{\partial}{\partial \dot u^\beta}{\mathcal L}={\mathcal L}
\\ &   & u^\beta\frac{\partial}{\partial \dot u^\beta}{\mathcal L}=0.
\end{eqnarray*}

Для пошуку самозмінних варіяційних рівнянь використовуємо два мірила \cdash--- мірило варіяційности і мірило самозмінности.

\section{Критерій варіяційности}
Розглядатимемо компоненти варіяційного рівняння~(\ref{matsy:fe}),
як компоненти ось якої диференційної один-форми:
\begin{equation}
\se=\sE_id\x^i.
\label{matsy:e}
\end{equation}

Для довільного $s\in\BBN$ нехай $\Omega_s(Q)$ означає алгебру
 диференційних форм на многовиді $T^sQ=\{\x^i,\sv^i, \sv'^i, \sv''^i,\dots, \sv_{(s-\SSS1)}^i\}$.
Нагадаємо поняття {\it упохіднення} в степенованих алгебрах,
наділених узагальненими переміжними співвідношеннями, якою і є
алгебра $\Omega_s(Q)$. Якийсь собі ділач $D$ зветься упохідненням
степеня $q$, якщо для будь-якої диференційної форми $\varpi$
степеня $p$ і будь-якої иньшої диференційної форми $w$
справджується співвідношення $D(\varpi \wedge w) = D(\varpi)
\wedge w +(-1)^{pq} \varpi \wedge D(w)$. Можна розвинути деяке
числення в алгебрі $\Omega_s(Q)$ шляхом впровадження ділачів, \cdash---
зовнішнього диференціяла $d$ і повної (або ж
формальної \glqq часової\,\grqq{}) похідної $D_t$, \cdash--- ось за якими приписами:
\[
df =\frac{\partial f}{\partial \x^i} d\x^i + \sum_r\frac{\partial
f}{\partial \sv_{(r)}^i} d\sv_{(r)}^i, \quad {d_v}^2 = 0\,;
\]
\[
D_tf=\frac{\partial f}{\partial t}+ \sv^i \frac{\partial f}{\partial
\x^i}+ \sum_r\sv_{(r+\SSS1)}^i \frac{\partial f}{\partial \sv_{(r)}^i}, \quad
D_td = dD_t.
\]
Для того, щоб запроваджене вище
означення стало повним, необхідно вимагати аби $d$ було
упохідненням степеня $1$, тоді як $D_t$ щоб було упохідненням
степеня $0$.  Для подальших рахунків ми потребуватимемо ще одного
ділача, \cdash--- упохіднення степеня $0$, \cdash--- що його позначимо $\iota$, і якого
означимо за посередництвом дії на функції та один-форми (які
разом локально породжують алгебру $\Omega_s(Q)$), таким чином:
\[
\iota f = 0,\quad \iota d\x^i = 0, \quad\iota d\sv^i = d\x^i,\quad
\iota d\sv_{(r)}^i = (r+1)\, d\sv_{(r-\SSS1)}^i.
\]

Нехай тепер ділач $\deg$ вимірює степінь диференційної форми. Нове
поняття {\em ляґранжевого диференціялу} $\delta$
впроваджується через свою дію на елементи з $\Omega_s(Q)$:
\[
\delta = \left(\deg + \sum_r\frac{(-1)^r}{r!}D_t{}^r \iota^r \right)d\,.
\]Оператор $\delta$ має властивість $\delta^2=0$.
Для
диференційно-геометричних об'єктів (\ref{matsy:L}) і
(\ref{matsy:e}) маємо співвідношення:
\begin{equation}
\se=\delta L.
\label{matsy:edelta}
\end{equation}
Тепер критерій того, що довільна сім\uapя виразів $\big\{\sE_i\big\}$
у формулі (\ref{matsy:e})  є лівою частиною варіяційного рівняння для деякого ляґранжіяну,
записується наступним чином~\cite{matsyuk:Tulczyjew}:
\begin{equation}
\delta \se = 0\,.
\label{matsy:crit}
\end{equation}

Правдомірі~(\ref{matsy:crit}) можна надати координатного виразу~\cite{matsyuk:Lawruk,matsyuk:MatMet1981}:
\[
\delta \se=\sum_{s=0}^r
\left(\dfrac{\partial\E_i}{\partial
\sv_{s-1}^j}-\sum_{k=s}^r(-1)^k\dfrac{k!}{(k-s)!s!}D_t^{k-s}\dfrac{\partial\E_j}{\partial
\sv_{k-1}^i}\right)d\sv_{s-1}^j\wedge d\x^i=0\,,
\]
звідкіль випливає система диференційних рівнянь з частковими похідними
\begin{subequations}\label{matsyuk:2}
\begin{align}
\label{matsyuk:2.1}
 \dfrac{\partial\E_i}{\partial \x^j}
 - \dfrac{\partial\E_j}{\partial \x^i}
+\sum_{k=0}^r(-1)^k D_t^k
\left(
\dfrac{\partial\E_i}{\partial \sv_{k-1}^j}
-\dfrac{\partial\E_j}{\partial \sv_{k-1}^i}
\right) &=0\,;&
 \\
\label{matsyuk:2.2}
\dfrac{\partial\E_i}{\partial
\sv_{s-1}^j}-\sum_{k=s}^r(-1)^k\dfrac{k!}{(k-s)!s!}D_t^{k-s}\dfrac{\partial\E_j}{\partial
\sv_{k-1}^i}&=0\,,& 1\leqslant s\leqslant r\,.
\end{align}
\end{subequations}
Записана тут система рівнянь є рівнозначною з наступною (яка отримана  із системи~(\ref{matsyuk:2.2})
поширенням обсягу зміни букви $s$ аж до залучення значення $s=0$):
\begin{align}\label{matsyuk:criterion}
\dfrac{\partial\E_i}{\partial
\sv_{s-1}^j}-\sum_{k=s}^r(-1)^k\dfrac{k!}{(k-s)!s!}D_t^{k-s}\dfrac{\partial\E_j}{\partial
\sv_{k-1}^i}&=0& 0\leqslant s\leqslant r\,.
\end{align}

\begin{proof}[Доведення]
Ускіснення виразу~(\ref{matsyuk:criterion}) при $s=0$ дає рівняння~(\ref{matsyuk:2.1}).

Навпаки, у рівнянні~(\ref{matsyuk:2.1}) відділімо доданок, який відповідає
$k=0$:
 \begin{equation*}%\label{matsyuk:2.1.3}
 2\,\dfrac{\partial\E_i}{\partial \x^j}
 - 2\,\dfrac{\partial\E_j}{\partial \x^i}
+\sum_{k=1}^r(-1)^k D_t^k
\dfrac{\partial\E_i}{\partial \sv_{k-1}^j}
-\sum_{k=1}^r(-1)^k D_t^k\dfrac{\partial\E_j}{\partial \sv_{k-1}^i}
 =0\,.
\end{equation*}
Під першим знаком суми замінімо
$\dfrac{\partial\E_i}{\partial \sv_{k-1}^j}$ його виразом з рівняння~(\ref{matsyuk:2.2}):
\[
\sum_{k=1}^r(-1)^k D_t^k \dfrac{\partial\E_i}{\partial \sv_{k-1}^j}
=\sum_{k=1}^r(-1)^k D_t^k
\sum_{s=k}^r(-1)^s\dfrac{s!}{(s-k)!k!}D_t^{s-k}\dfrac{\partial\E_j}{\partial
\sv_{s-1}^i}\,.
\]
Перемінімо порядок сумування:
$\sum_{k=1}^r\sum_{s=k}^r=\sum_{\substack{ s,k=1
\\ s\geqslant k}}^r=\sum_{s=1}^r\sum_{k=1}^s$.
Підрахуймо суму за  $k$:
\[
\sum_{k=1}^s(-1)^k\dfrac{s!}{(s-k)!k!}=\sum_{k=0}^s(-1)^k \binom s k -
\binom s 0 =0-1=-1\,.
\]
Врешті рівняння~(\ref{matsyuk:2.1})
переходить у
 \[ 2\,\dfrac{\partial\E_i}{\partial \x^j} -
2\,\dfrac{\partial\E_j}{\partial \x^i} -\sum_{k=1}^r(-1)^k D_t^k
 \dfrac{\partial\E_j}{\partial \sv_{k-1}^i} -\sum_{k=1}^r(-1)^k
D_t^k\dfrac{\partial\E_j}{\partial \sv_{k-1}^i} =0\,, \]
яке збігається з подвоєним рівнянням~(\ref{matsyuk:criterion}) при $s=0$.
\end{proof}
Правдоміра~(\ref{matsyuk:criterion}) отримувалася ріжними авторами. Щодо огляду письмен з цього приводу відсилаємо до книги~\cite{matsyuk:OKrup}.

Зосередимося на рівняннях третього порядку. Є очевидним, що вираз, який відповідає лівій частині такого рівняння, має афінний вигляд щодо найстарших похідних. Застосуємо деякі звичні векторні позначки: крапка долі означає згортку рядочка з наступним стовпцем, а часом також позначатиме матричне множення між матрицею і наступним стовпцем.
Частково розв\uapязуючи систему рівнянь~(\ref{matsyuk:criterion}) і по можливості спрощуючи, можна дійти висновку, що загальний вигляд рівнянь Ойлєра\cdash--~Пуасона \emph{третього порядку} є ось:
\begin{equation}\label{matsyuk:hamspin5}
\A{\,\bkey.\,}\V^{{{\prime}}{{\prime}}}{\, +
\,}(\V^{{\prime}}{\!\bkey.\,}{\bpartial}_{\V})\,
\A{\,\bkey.\,}\V^{{\prime}}{\, + \,}\B{\,\bkey.\,}\V^{{\prime}}{\, +
\,}\CC\, = \,\Zero\,,
\end{equation}
де скісна матриця
$\A$, симетрична матриця $\B$ і стовпець $\CC$, -- усі залежать од $t$, $\x^i$,
 $\sv^i$ та задовольняють ось яку систему диференційних рівнянь з частковими похідними:
\begin{equation}\label{matsyuk:hamspin6}
{\renewcommand{\arraystretch}{1.7}
\begin{gathered}
        \partial_{_{_{_{{{\sv}}}}}}{\!}_{[i}{}{{\sA}}_{jl]}=0
\\
         2\,{{\sB}}_{[ij]}-3\,{\bf D_{_{\bmat 1}}}{\kern.01667em}{{\sA}}_{ij}=0
\\
      2\,\partial_{_{_{_{{{\sv}}}}}}{\!}_{[i}{}{{\sB}}_{j]\,l}
               -4\,\partial_{_{_{_{{{\x}}}}}}{\!}_{[i}{}{{\sA}}_{j]\,l}
               +{\partial_{_{_{_{{{\x}}}}}}{\!}_{l}}{\,}{{\sA}}_{ij}
   +2\,{\bf D_{_{\bmat 1}}}{\kern.01667em}{\partial_{_{_{_{{{\sv}}}}}}{\!}_{l}}{\,}{{\sA}}_{ij}=0
\\
       {\partial_{_{_{_{{{\sv}}}}}}{\!}_{(i}}{}{{\sC}}_{j)}
               -{\bf D_{_{\bmat 1}}}{\kern.01667em}{{\sB}}_{(ij)}=0
\\
     2\,{\partial_{_{_{_{{{\sv}}}}}}{\!}_{l}}{\,}\partial_{_{_{_{{{\sv}}}}}}{\!}_{[i}{}{{\sC}}_{j]}
           -4\,\partial_{_{_{_{{{\x}}}}}}{\!}_{[i}{}{{\sB}}_{j]\,l}
           +{{\bf D_{_{\bmat 1}}}}^{2}{\,}{\partial_{_{_{_{{{\sv}}}}}}{\!}_{l}}{\,}{{\sA}}_{ij}
 +6\,{\bf D_{_{\bmat 1}}}{\kern.0334em}\partial_{_{_{_{{{\x}}}}}}{\!}_{[i}{}{{\sA}}_{jl]}=0
\\
4\,\partial_{_{_{_{{{\x}}}}}}{\!}_{[i}{}{{\sC}}_{j]} -2\,{\bf D_{_{\bmat
           1}}}{\kern.0334em}\partial_{_{_{_{{{\sv}}}}}}{\!}_{[i}{}{{\sC}}_{j]} -{{\bf D_{_{\bmat 1}}}}^{3}{\,}{{\sA}}_{ij}=0\,.
\end{gathered}}
\end{equation}
Тут диференційний ділач $\bf D_{_{\bmat 1}}$ є обтятим до найнижчого порядку ділачем повної похідної для змінних~$\x^i$,
\[
{\bf D_{_{\bmat 1}}}=\partial_{t}{\,+\,}\V{\,\bkey.\,}{\bpartial}_{\X}\,.
\]
\section{Критерій варіяційности в одноманітних співрядних.}
Подібним чином виражається правдоміра варіяційности для рівняння третього порядку в одноманітних співрядних.
{\let\A=\bA
\let\B=\bB
\let\CC=\bC
\let\sA=\rA
\let\sB=\rB
\let\sC=\rC
\let\sv=\iu
\let\x=\ix
\let\V=\bU
\let\X=\bX
Варіяційне рівняння~(\ref{matsy:CalE}) прибирає конкретнішого вигляду
\begin{equation}\label{matsyuk:hamspin5hom}
\A{\,\bkey.\,}\V^{{{\prime}}{{\prime}}}{\, +
\,}(\V^{{\prime}}{\!\bkey.\,}{\bpartial}_{\V})\,
\A{\,\bkey.\,}\V^{{\prime}}{\, + \,}\B{\,\bkey.\,}\V^{{\prime}}{\, +
\,}\CC\, = \,\Zero\,,
\end{equation}
де скісна матриця
$\A$, симетрична матриця $\B$ і стовпець $\CC$, \cdash--- усі залежать від  $\x^\alpha $ та
 $\sv^\alpha$, а ще й задовольняють таку систему рівнянь з частковими похідними:
\begin{equation}\label{matsyuk:hamspin6hom}
{\renewcommand{\arraystretch}{1.7}
\begin{gathered}
       \partial_{_{_{_{{{\sv}}}}}}{\!}_{[\alpha }{}{{\sA}}_{\beta  \lambda ]}=0
\\
       2\,{{\sB}}_{[\alpha \beta  ]}-3\,{{\boldsymbol{\mathcal D}}_{_{\bmat 1}}}{\kern.01667em}{{\sA}}_{\alpha \beta  }=0
\\
    2\,\partial_{_{_{_{{{\sv}}}}}}{\!}_{[\alpha }{}{{\sB}}_{\beta  ]\,\lambda }
               -4\,\partial_{_{_{_{{{\x}}}}}}{\!}_{[\alpha }{}{{\sA}}_{\beta  ]\,\lambda }
               +{\partial_{_{_{_{{{\x}}}}}}{\!}_{\lambda }}{\,}{{\sA}}_{\alpha \beta  }
   +2\,{{\boldsymbol{\mathcal D}}_{_{\bmat 1}}}{\kern.01667em}{\partial_{_{_{_{{{\sv}}}}}}{\!}_{\lambda }}{\,}{{\sA}}_{\alpha \beta  }=0
\\
     {\partial_{_{_{_{{{\sv}}}}}}{\!}_{(\alpha }}{}{{\sC}}_{\beta  )}
               -{{\boldsymbol{\mathcal D}}_{_{\bmat 1}}}{\kern.01667em}{{\sB}}_{(\alpha \beta  )}=0
\\
    2\,{\partial_{_{_{_{{{\sv}}}}}}{\!}_{\lambda }}{\,}\partial_{_{_{_{{{\sv}}}}}}{\!}_{[\alpha }{}{{\sC}}_{\beta  ]}
           -4\,\partial_{_{_{_{{{\x}}}}}}{\!}_{[\alpha }{}{{\sB}}_{\beta  ]\,\lambda }
           +{{{\boldsymbol{\mathcal D}}_{_{\bmat 1}}}}^{2}{\,}{\partial_{_{_{_{{{\sv}}}}}}{\!}_{\lambda }}{\,}{{\sA}}_{\alpha \beta  }
 +6\,{{\boldsymbol{\mathcal D}}_{_{\bmat 1}}}{\kern.0334em}\partial_{_{_{_{{{\x}}}}}}{\!}_{[\alpha }{}{{\sA}}_{\beta  \lambda ]}=0
\\
4\,\partial_{_{_{_{{{\x}}}}}}{\!}_{[\alpha }{}{{\sC}}_{\beta  ]} -2\,{{\boldsymbol{\mathcal D}}_{_{\bmat
           1}}}{\kern.0334em}\partial_{_{_{_{{{\sv}}}}}}{\!}_{[\alpha }{}{{\sC}}_{\beta  ]} -{{{\boldsymbol{\mathcal D}}_{_{\bmat 1}}}}^{3}{\,}{{\sA}}_{\alpha \beta  }=0\,.
\end{gathered}}
\end{equation}
Тут диференційний ділач
 ${\boldsymbol{\mathcal D}}_{_{\bmat 1}}$ є обтятим до найнижчого порядку ділачем повної похідної для змінних~$\x^\alpha$,
\[
{{\boldsymbol{\mathcal D}}_{_{\bmat 1}}}=
\V{\,\bkey.\,}{\bpartial}_{\X}\,.
\]
}

Тепер вкажемо, як вирахувати матриці $\bA$, $\bB$ та стовпець $\bC$, маючи в розпорядженні матриці $\A$, $\B$, разом із стовпцем $\CC$.
\begin{lema}\label{matsyuk:bbMatrices}
%\begin{enumerate}
%  \item
Нехай у рівнянні~(\ref{matsyuk:hamspin5hom})
    \begin{align*}
        \bA&=\begin{pmatrix}
              0 & -\ba \\
              \ba^T & \mathbb A \\
            \end{pmatrix}\,,\\
        \bB&=\begin{pmatrix}
               \rB_{\so\so} & \tilde{\bb} \\
               \bb & \mathbb B\\
             \end{pmatrix}\,,\\
        \bC&=\begin{pmatrix}
               \rC_{\so}  \\
               {\scriptstyle\mathbb C}\\
             \end{pmatrix}\,.
    \end{align*}
Нехай так само у виразі~(\ref{matsy:CalE})
\begin{equation*}
    \{\mathcal E_{\alpha}\} =\begin{pmatrix}
                               \mathcal E_{\so} \\
                               \mathbb E \\
                             \end{pmatrix}\,.
\end{equation*}

Тоді виражаємо величини рівняння~(\ref{matsyuk:hamspin5hom}) у величинах рівняння~(\ref{matsyuk:hamspin5}):
\begin{align*}
    \mathbb A &= \frac{1}{\dot t^{\strut2}}\,\A\circ p^\so\,,\\
    \mathbb B &= \frac{1}{\dot t^{\strut}}\,\B\circ p^\so\,,\\
   {\scriptstyle\mathbb C} & = \dot t\, \CC\circ p^\so\,,\\
\end{align*}
де матриця $\rB$ є симетричною, і, окрім цього,
\begin{equation*}
   \mathbb A\bdot\V+\ba=0\,,\qquad \mathbb B\bdot\V+\bb=0\,,\qquad \rB_{\so\so}+\bb\bdot\V=0\,,\qquad \rC_{\so}+{\scriptstyle\mathbb C}\bdot\V=0\,;
\end{equation*}
\begin{equation*}
\mathbb E= \dot t^3\left[\A{\,\bkey.\,}\V^{{{\prime}}{{\prime}}}{\, +
\,}(\V^{{\prime}}{\!\bkey.\,}{\bpartial}_{\V})\,
\A{\,\bkey.\,}\V^{{\prime}}{\, + \frac{1}{\dot t^{\strut}}\,}\B{\,\bkey.\,}\V^{{\prime}}{\, +
\,}\frac{1}{\dot t^{\strut3}}\,\CC\right]\,.
\end{equation*}
B укладі~(\ref{matsy:CalE}) виконується умова Вейєрштраса~\cite{matsyuk:Logan}
\begin{equation*}
    \bU\bdot\boldsymbol{\mathcal E}=0\,.
\end{equation*}
 \end{lema}

\begin{zauv}
Скісні властивості матриць, \cdash--- матриці~$\bA$ в укладі~(\ref{matsyuk:hamspin5hom}) і матриці~$\A$ в укладі~(\ref{matsyuk:hamspin5}), \cdash--- накладають гострі обмеження на саме існування варіяційного рівняння третього порядку в нижчих вимірах:
\begin{enumerate}
  \item Не існує варіяційного рівняння третього порядку в одновимірному просторі (значок $\alpha$ прибирає одного"=єдиного значення, $\alpha=0$);
  \item Не існує безвідмірного (себто \glqq параметрично\cdash--~інваріянтного\grqq{}) варіяційного рівняння на площині (значок $i$ прибирає одного"=єдиного значення, $i=1$).
\end{enumerate}
\end{zauv}
\section{Евклідівська незмінність}
Ґраф інтеґральної доріжки $t\mapsto\x^i(t)$, $i=1,\dots n$, варіяційного рівняння~(\ref{matsy:fe}) можна продовжити до перекрою $t\mapsto(t,\x^i(t),\sv^i(t),\sv'\msp^i(t),\sv''\msp^i(t))$ в'язки струменів $J^3(\R,\R^n) \to\R$, інтеґрального щодо векторної диференційної один-форми у змінній $t$,
\begin{equation}\label{matsyuk:UJP23}
\begin{aligned}
e&=\alpha\otimes dt&&\text{\big(гляди~(\ref{matsy:e})\big)}\\
&=\E_i d\x^i\otimes dt\,,&&
\end{aligned}
\end{equation}
заданої на просторі $J^3(\R,\R^n)$ струменів перекроїв прямокутного добутку $\R\times\R^n$.

Поряд з диференційною формою~(\ref{matsyuk:UJP23}) зручно впровадити ще й так званий її \emph{лепажівський еквівалент}, що його чинники не залежать од похідних третього порядку:
\begin{align}
\label{matsyuk:UJP31}
\epsilon=\sA_{ij} d\x^i\otimes d\sv^{\prime}{}^j + &\, \sk_i d\x^i \otimes dt, &\\
&\,\K = (\V^{{\prime}}{\!\bkey.\,}{\bpartial}_{\V})\,
\A\,\bkey.\,\V^{{\prime}}{\, + \,}\B\,\bkey.\,\V^{{\prime}}{\, +
\,}\CC\,.\label{matsyuk:K}
\end{align}
Про цю векторно"=значну диференційну один"=форму, що набирає значень з простору
$T^*\R^n$, можна думати, як про деяку інтерпретацію поняття \emph{лепажівської форми}, альтернативну до викладеної в книзі~\cite{matsyuk:OKrup}.
Оскільки ми зацікавлені в голономних доріжках,
як звичайно, вважатимемо векторно"=значні диференційні один"=форми (\ref{matsyuk:UJP31}) і (\ref{matsyuk:UJP23}) рівнозначними в стосунку до модуля торкання на многовиді~$J^3(\R,\R^n)$,
\[
\epsilon-e =\sA_{ij}d\x^i \otimes   \theta_{\bmat3}^j\,,
\]
де векторно"=значні один"=форми торкання
\begin{equation}
\label{matsyuk:hamspin8}
\boldsymbol{\theta_1}=d\X-\V dt\,,\quad\boldsymbol{\theta_2}= d \V-\V' dt\,,
\quad \boldsymbol{\theta_3}=d\V'-\V'' dt
\end{equation}
породжують модуль торкання на многовиді~$J^3(\R,\R^n)$.

Щоби враз охопити, як справді евклідівський, так і лже"=евклідівський випадки, узгоднімо деякі позначки.

Буквою $\eta$ позначім знак $+$ чи $-$ компоненти $g_{\so\,\so}$ звичаєвого діагонального метричного тензора.
Серединною крапкою позначім операцію скалярного добутку поміж матрицями, які виражають тензори, або ж поміж вервечками, які виражають вектори, \cdash--- стосовно звичаєвого (лже)ев\-клі\-дів\-сько\-го метричного тензора. Таким чином, скалярний добуток є нічим иньшим, як просто згорткою, до якої залучений метричний тензор. Твірник  $X$ (лже)евклідівських перетворень у тривимірному просторі можна параметризувати деякою скісною матрицею $\W$ і деяким вектором $\bP$:
\begin{eqnarray}
 X&=&
-\,(\bP\bcdot{\X})\,{\partial_t}+\eta\,t\,\bP\,\bkey.\,\bpartial_{\X}
+\W\bcdot({\X}\wedge\bpartial_{\X})
\nonumber\\ &&{}+\eta\,\bP\,\bkey.\,\bpartial_{\V}
+(\bP\bcdot \V)\,\V\,\bkey.\,\bpartial_{\V}+\W\bcdot(\V\wedge\bpartial_{\V})\nonumber
\\ \nonumber &&{}+2\,(\bP\bcdot \V)\,\V'\bkey.\,\bpartial{_{\V'}}+(\bP\bcdot \V')\,\V\,\bmat.\,\bpartial{_{\V'}}
+\W\bcdot(\V'\wedge\bpartial{_{\V'}})\,.
\end{eqnarray}
Є можливим вкласти поняття симетрії рівняння~(\ref{matsyuk:hamspin5})
в загальні рамці науки про незмінність зовнішньої диференційної системи. Система, про яку нам йдеться, породжена векторно"=значною пфафівською формою $\bmat\epsilon$ з укладу~(\ref{matsyuk:UJP31}) та векторно"=значними диференційними формами торкання
$\boldsymbol{\theta_1}$ і
$\boldsymbol{\theta_2}$ з укладу~(\ref{matsyuk:hamspin8}).
Нехай $X(\bmat\epsilon)$ означає похідну Лі від векторно"=значної диференційної форми $\bmat\epsilon$ уздовж векторного поля~$X$. Умова незмінности полягає в тім, що мали"~б існувати деякі такі матриці ${\bf\Phi}$, ${\bf\Xi}$, і ${\bf\Pi}$, залежні од
 $\V$ і $\V'$, що
\begin{equation}\label{matsyuk:hamspin9}
X(\bmat\epsilon)={\bf\Phi}\,\bmat.\,\bmat\epsilon+{\bf\Xi}\,.\,(d\X-\V d t)
+{\bf\Pi}\,\bmat.\,(d\V-\V' d t).
\end{equation}
Також припускатимемо, що
 $\A$ і $\K$ in~(\ref{matsyuk:UJP31}) не залежать
ані від $t$ ані від $\X$.
\begin{prop}\label{matsyuk:noexistence}
В чотиривимірному (лже)евклідівському просторі не існує самозмінних варіяційних рівнянь третього порядку
\end{prop}

\begin{proof}[Доведення]
Умова самозмінности~(\ref{matsyuk:hamspin9}) розпадається на окремі тотожності відповідно до прирівнювання чинників при диференціялах $d\V'$, $d\V$,$d\X$, $dt$:
\begin{gather}\label{diss11.1}
    \left[\bP\,\bkey.\,\bpartial_{\V}
+(\bP\bcdot \V)\,\V\,\bkey.\,\bpartial_{\V}+\W\bcdot(\V\wedge\bpartial_{\V}\right]\A
+2\,(\bP\bcdot \V)\,\A+
\A\bkey.\,\V\otimes\bP-\A\,\bkey.\,\W=\bf\Phi\,\bkey.\,\A
\\
2\,(\A\bdot\V')\otimes\bP+(\bP\bdot\V')\,\A=\bf\Pi
\\
-\K\otimes\bP=\bf\Xi
\\
X\K=\bf\Phi\bdot\K-\bf\Xi\bdot\V-\bf\Pi\bdot\V'
\end{gather}
Скісна матриця обшару $3$ завжди є виродженою: якщо позначимо
\begin{equation*}\label{diss12.9}
   \a\overset{\text{def}}=\ast\,\A\,,
\end{equation*}
то матимемо $\A\bdot\a=0$. Згорнімо, коли так, рівняння~(\ref{diss11.1}) зі стовпцем $\a$ і розщепімо за параметрами $\bP$ та $\w\overset{\mathrm{def}}=\ast\,\W\,$:
\begin{align}\label{diss13.14}
    \fa\times(\bP\bkey.\,\bpartial_{\V})\;\fa+(\bP\bcdot\V)\,\fa\times(\V\bkey.\,\bpartial_{\V})\,\fa
    -(\bP\bcdot\fa)\;\fa\times\V=0
\\
    \fa\times[\w\,\V\,\bpartial_{\V}]\fa-\fa\times(\w\times\fa)=0\,.
\label{diss13.15}
\end{align}
Покладімо в рівнянні~(\ref{diss13.14}) $\bP=\w\times\V$ і використаймо~(\ref{diss13.15}):
\begin{equation*}
    \fa\times(\w\times\fa) - [\w\,\V\,\fa]\fa\times\V=0\,.
\end{equation*}
Згорнімо зі стовпцем $\w$:
\begin{equation}\label{matsyuk:diss15'}
    (\fa\times\w)^{\SSS\boldsymbol2}+[\fa\,\V\,\w]^{\SSS\boldsymbol2}=0\,.
\end{equation}
Величина $(\fa\times\w)^{\SSS\boldsymbol2}=\fa^{\SSS\boldsymbol2}\w^{\SSS\boldsymbol2}-(\fa\bcdot\w)^{\SSS\boldsymbol2}$
є додатною, якщо сиґнатура метрики дорівнює $\pm3$. В решті випадків, користаючи з того, що $(\fa\times\w)\bcdot\w=0$, завжди можна вибрати стовпець $\w$ таким чином, щоб вектор $\fa\times\w$ не стримів в уявний бік, $(\fa\times\w)^{\SSS\boldsymbol2}\geq0$. Тому із співвідношення~(\ref{matsyuk:diss15'}) випливає, що $(\fa\times\w)^{\SSS\boldsymbol2}=0$. З огляду на решту довільности у виборі~$\w$, повинно бути~$\fa=0$.
\end{proof}
\section{Варіяційні рівняння третього порядку для вільної дзиґи.}
Як видно із Речі",\ref{matsyuk:noexistence}, спроби збудувати пуанкаре"--~самозмінне варіяційне рівняння третього порядку в чотиривимірному світі приречені на невдачу. Ми обминаємо цю трудність упровадженням до шуканого рівняння додаткового вектор"--~параметра~$\bs$, який перетворювався б за виказом дії групи Лоренца. Потрібно, поруч із цим, стежити, аби не порушувалася будова рівняння Ойлєра"--~Пуасона~(\ref{matsyuk:hamspin5}) разом з умовами~(\ref{matsyuk:hamspin6}).

Нехай, отже, вираз~(\ref{matsyuk:K}) не містить явної залежности від змінних $t$ і $\X$, а, зате, величини $\fa=\ast\,\A$, $\B$ та $\CC$ містять залежність ще й від чотири"--~вектора $\bs=\big(s^{\so},\,\bbs\big)$. Твірник перетворень Лоренца повинен містити часть, яка діятиме на параметри $s^{\so}$ та $\bbs$ ($\eta=1$):
\begin{eqnarray*}
 X&=&
-\,(\bP\bcdot{\bbs})\,{\partial_{_{{\scriptstyle s}^{\scriptscriptstyle 0}}}}+\,s^{\so}\,\bP\,\bkey.\,\bpartial_{\bbss}
+[\w\,\bbs\,\bpartial_{\bbss}]\\&&
-\,(\bP\bcdot{\X})\,{\partial_t}+\,t\,\bP\,\bkey.\,\bpartial_{\X}
+[\w\,\X\,\bpartial_{\X}]
\\ &&{}+\,\bP\,\bkey.\,\bpartial_{\V}
+(\bP\bcdot \V)\,\V\,\bkey.\,\bpartial_{\V}+[\w\,\V\,\bpartial_{\V}]
\\  &&{}+2\,(\bP\bcdot \V)\,\V'\bkey.\,\bpartial{_{\V'}}+(\bP\bcdot \V')\,\V\,\bmat.\,\bpartial{_{\V'}}
+[\w\,\V'\,\bpartial_{\V'}]\,.
\end{eqnarray*}
Умова самозмінности~(\ref{matsyuk:hamspin9}) розпадається на окремі тотожності, які утворюються від прирівнювання чинників при диференціялах $d\V'$, $d\V$, $d\X$ та $dt$:
\begin{multline}\label{matsyuk:diss16.2}
  \left(\bP\bcdot\bbs\;\partial_{_{{\scriptstyle s}^0}}
  -s^{\so}\,\bP\bdot\bpartial_{\bbss}-[\w\,\bbs\,\bpartial_{\bbss}]
-\bP\bdot\bpartial_{\V}-(\bP\bcdot\V)\,\V\bdot\bpartial_{\V}-[\w\,\V\,\bpartial_{\V}]\right)\,\fa\times d\V'
\\
-2\,(\bP\bcdot\V)\,\fa\times d\V'-\fa\times\V\,(\bP\bcdot d\V')-\fa\times(\w\times d\V')
=-\,{\bf\Phi}\bdot(\fa\times d\V')\,;
\end{multline}
\begin{equation}\label{matsyuk:diss16.3}
    -\,2\,\fa\times\V'\,(\bP\bcdot d\V)-(\bP\bcdot\V')\,\fa\times d\V={\bf\Pi}\bdot d\V\,;
\end{equation}
\begin{equation}\label{matsyuk:diss16.4}
   -\K\,(\bP\cdot d\X)={\bf \Xi}\bdot d\X
\end{equation}
\begin{equation}\label{matsyuk:diss16.5}
    X\K={\bf\Phi}\bdot\K-{\bf\Xi}\bdot\V-{\bf\Pi}\bdot\V'
\end{equation}
Оскільки скісна матриця $\A$ є виродженою, рівнянні~(\ref{matsyuk:diss16.2}) містить деяке співвідношення, до якого не входить невизначений чинник~$\bf \Phi$. Ось, покладімо у рівнянні~(\ref{matsyuk:diss16.2}) $d\V'=\fa$. Права часть рівняння щезне,
так що отримаємо два співвідношення з параметрами, відповідно, $\bP$ та $\w$:
 \begin{align}\label{matsyuk:diss16.6}
    \fa\times\left[-\,\bP\bcdot\bbs\,\partial_{_{{\scriptstyle s}^0}}+s^{\so}\,\bP\bdot\bpartial_{\bbss}+\bP\bdot\bpartial_{\V}+(\bP\bcdot\V)\,\V\bdot\bpartial_{\V}\right]\fa -(\bP\bcdot\fa)\,\fa\times\V&=0\,,\\
\fa\times\left([\w\,\bbs\,\bpartial_{\bbss}]+[\w\,\V\,\bpartial_{\V}]\right)\fa-\fa\times(\w\times\fa)&=0\,.\label{matsyuk:diss16.7}
 \end{align}
В укладі~(\ref{matsyuk:diss16.6}) покладемо один раз $\bP=s^{\so}\,\w\times\V$, а за другим разом $\bP=\w\times\bbs$, та й додамо отримані вирази. Згідно з укладом~(\ref{matsyuk:diss16.7}) одержуємо:
\begin{equation}\label{matsyuk:diss16.8}
\begin{split}
-\,s^{\so}\,[\w\,\V\,\bbs]\,\fa\times\partial_{_{{\scriptstyle s}^0}}\fa+s_{\so}{}^2\fa\times[\w\,\V\,\bpartial_{\bbss}]\,\fa
-\,s^{\so}\,[\w\,\V\,\fa]\,\fa\times\V+\fa\times[\w\,\bbs\,\bpartial_{\V}]\,\fa \hphantom{(}&  \\
+\,[\w\,\bbs\,\V]\,\fa\times(\V\bdot\bpartial_{\V})\,\fa-[\w\,\bbs\,\fa]\,\fa\times\V+s^{\so}\fa\times(\w\times\fa)                                            & =0\,.
\end{split}
\end{equation}
Запровадьмо позначку
\begin{equation*}
    \ff=\bbs-s^{\so}\V\,.
\end{equation*}
Покладімо в~(\ref{matsyuk:diss16.7}) і у~(\ref{matsyuk:diss16.8}) $\w=-\ff$. Помножім~(\ref{matsyuk:diss16.7}) на $-s^{\so}$ і збудуймо півсуму з~(\ref{matsyuk:diss16.8}):
\begin{equation}\label{matsyuk:diss16.9}
    -[\ff\,\V\,\fa]\,\fa\times\V+\fa\times(\ff\times\fa)=0\,.
\end{equation}
Тепер помножім~(\ref{matsyuk:diss16.9}) скалярно на вектор $\ff$:
\begin{equation}\label{matsyuk:diss16.10}
   [\fa\,\V\,\ff]^2+(\fa\times\ff)^{\boldsymbol2}=0\,.
\end{equation}
Якщо $(\fa\times\ff)^{\boldsymbol2}=0$, при довільних $s^{\so}$, $\bbs$, $\V$, то вектор $\ff$ паралельний до вектора~$\fa$:
\begin{equation}\label{matsyuk:diss16.11}
    \fa=\mathfrak a(\sv^\alpha)\,\ff\,.
\end{equation}
Розв'язка~(\ref{matsyuk:diss16.11}) задовольняє рівняння~(\ref{matsyuk:diss16.6}, \ref{matsyuk:diss16.7}).

У тотожність~(\ref{matsyuk:diss16.5}) можна підставити неозначені чинники $\bf\Xi$ та $\Pi$
 із ~(\ref{matsyuk:diss16.4}) та ~(\ref{matsyuk:diss16.3}):
 \begin{equation}\label{matsyuk:diss16.12}
    X\K={\bf\Phi}\bdot\K+(\bP\bcdot\V)\,\K+3\,(\bP\bcdot\V')\,\fa\times\V'\,.
 \end{equation}
Позначімо $\G =\begin{pmatrix}
                       g_{ij} \\
                     \end{pmatrix}
$ і запровадьмо матрицю
\begin{equation*}
    \mathbf W=-X\A-2(\bP\bcdot\V)\,\A-(\A\bdot\V)\otimes\bP+\w\otimes\fa-(\w\bcdot\fa)\G \,.
\end{equation*}
Тепер тотожність~(\ref{matsyuk:diss16.2}) запишеться у скороченому вигляді:
\begin{equation}\label{matsyuk:diss16.13}
  {\bf\Phi}\times\fa=\mathbf W\,,
\end{equation}
де векторний добуток поміж діадиком~${\bf\Phi}$ та вектором~$\fa$ запроваджується взором
\begin{equation*}
    ({\bf \Phi}\times\fa)\bdot \n={\bf \Phi}\bdot(\fa\times \n)\,
\end{equation*}
при довільному векторі~$\n$.
Перемноживши~(\ref{matsyuk:diss16.13}) справа векторно на $\K$, з поміччю укладу
\begin{equation*}
    ({\bf\Phi}\times\fa)\times\K=({\bf\Phi}\bdot\K)\otimes\fa-(\fa\bdot\K)\,{\bf\Phi}
\end{equation*}
отримаємо, після підставлення~(\ref{matsyuk:diss16.12}),
\begin{equation}\label{matsyuk:diss16.14}
   \left[X\K-(\bP\bcdot\V)\,\K-3\,(\bP\bcdot\V')\,\fa\times\V'\right]\otimes\fa
   -\mathbf W\times\K-(\fa\bcdot\K)\,{\bf\Phi}=0\,.
\end{equation}
Тотожності (\ref{matsyuk:diss16.12}, \ref{matsyuk:diss16.13}) алгебрично рівнозначні з тотожністю~(\ref{matsyuk:diss16.14}), якщо тільки $\fa$ і $\K$ є такими, що $\fa\bcdot\K\ne 0$.
Підставмо (\ref{matsyuk:diss16.11}) у~(\ref{matsyuk:diss16.14}) і скористаймо з~(\ref{matsyuk:diss16.13}).
Отримуємо:
\begin{equation}
\label{matsyuk:diss16.15}
    \left[X\K-(\bP\bcdot\V)\,\K-3\,(\bP\bcdot\V')\,\fa\times\V'\right]\otimes\fa+\frac{\mathfrak a'}{\mathfrak a}\,\mathbf W\times\fa- \mathbf W\times\left({\B}\bdot\V'+\CC\right)=(\fa\bcdot\K)\,{\bf\Phi}\,.
\end{equation}

Нижче подані функції задовольняють систему рівнянь з частковими похідними~(\ref{matsyuk:diss16.15}) і, водночас, задовольняють також і умови~(\ref{matsyuk:hamspin6})
\begin{equation*}
    \fa=\left[(1+\V^{\boldsymbol2})(s_{\so}{}^2+\bbs^{\boldsymbol2})-(s^{\so}+\bbs\bcdot\V)^2\right]^{-3/2\,;}
\end{equation*}
\begin{equation*}
    \B=\mu\,\frac{(1+\V^{\boldsymbol2})\G - \V\otimes\V}{(1+\V^{\boldsymbol2})^{3/2}(s_{\so}{}^2+\bbs^{\boldsymbol2})^{3/2}}\,;
\end{equation*}
\begin{equation*}
    \CC=0\,.
\end{equation*}
Рівняння Ойлєра\cdash--~Пуасона прибирає вигляду:
\begin{multline}\label{matsyuk:UltimateNonhom}
\boldsymbol\sE
=\frac{\V''\times(\bbs-s^{\so}\V)}{\left[(1+\V^{\boldsymbol2})(s_{\so}{}^2+\bbs^{\boldsymbol2})
-(s^{\so}+\bbs\bcdot\V)^2\right]^{3/2}}
\\
-3\,\frac{(s_{\so}{}^2+\bbs^{\boldsymbol2})\,\V'\bcdot\V-(s^{\so}+\bbs\bcdot\V)\,\bbs\bcdot\V'}{\left[(1+\V^{\boldsymbol2})(s_{\so}{}^2+\bbs^{\boldsymbol2})-(s^{\so}+\bbs\bcdot\V)^2\right]^{5/2}}
   \,\V'\times(\bbs-s^{\so}\V)
\\
  +\frac{\mu}{(1+\V^{\boldsymbol2})^{3/2}(s_{\so}{}^2+\bbs^{\boldsymbol2})^{3/2}}\left[(1+\V^{\boldsymbol2})\V'
    -(\V'\bcdot\V)\,\V\right]=0\,.
\end{multline}
Послуговуючись взором~(\ref{matsyuk:hamspin5hom}) і Лемою на сторінці~\pageref{matsyuk:bbMatrices}, отримуємо рівняння~(\ref{matsyuk:26}).

Вираз у лівій часті рівняння~(\ref{matsyuk:UltimateNonhom}) є виразом Ойлєра\cdash--~Пуасона для кожної з ось якої сім\uapiї функцій Ляґранжа:
\begin{multline*}
    L_{(i)}=\frac{s_{\so}}{s_{\so}{}^2+\bbs^{\boldsymbol2}}\,\bcdot\,
    \frac{(s_{\so}{}^2+\n_{(i)}{}^{\boldsymbol2})(\sfs_i-s_{\so}\sv_i)
-\sfs_i(\n_{(i)}\bcdot\z_{(i)})}{(s_{\so}{}^2+\n_{(i)}{}^{\boldsymbol2})\z_{(i)}{}^{\boldsymbol2}-(\n_{(i)}\bcdot\z_{(i)})^2}\,\bcdot\,
    \frac{\left[\V'\;(\bbs-s_{\so}\V)\;\sfe_{(i)}\right]}{(\bbs-s_{\so}\V)^{\boldsymbol2}+(\bbs\times\V)^{\boldsymbol2}}
\\
-\frac{\mu}{(s_{\so}{}^2+\bbs^{\boldsymbol2})^{3/2}}\,\sqrt{1+\V^{\boldsymbol2}}\,,
\end{multline*}
де запроваджено позначки:
\begin{equation*}
    \n_{(i)}=\bbs-\sfs_i\sfe_{(i)}\,,\qquad\z_{(i)}=(\bbs-s_{\so}\V)-(\sfs_i-s_{\so}\sv_i)\sfe_{(i)}\,,
\end{equation*}
а вектори $\sfe_{(i)}$ утворюють базу в~$\mathds E^3$.

Відповідно до Речі",\ref{matsy:homogen}, із повищої сім\uapiї функцій Ляґранжа отримуємо для рівняння~(\ref{matsyuk:26}) ось яку сім\uapю функцій Ляґранжа~(\ref{matsy:L0}):
\begin{equation*}
    \mathcal L_{(\beta)}=
    \frac{
    \ast\,\boldsymbol{\dot u}\wedge\boldsymbol{u}\wedge\boldsymbol s\wedge\boldsymbol{e_{(\beta)}}
    }
    {\norm{\boldsymbol s}^2\norm{\boldsymbol s\wedge\boldsymbol u}}\,\bcdot\,
\frac{\boldsymbol s^{\boldsymbol2}u_\beta+(\boldsymbol s\bcdot\boldsymbol u)\,s_\beta}{(u_\beta\boldsymbol s-s_\beta\boldsymbol u)^{\boldsymbol2}-(\boldsymbol s\wedge\boldsymbol u)^{\boldsymbol2}}
    -\frac{\mu}{\norm{\boldsymbol s}^3}\norm{\boldsymbol u}\,,
\end{equation*}
де вектори $\boldsymbol e_{(\beta)}=\{\boldsymbol e_{\so},\,\sfe_{(i)}\}$ утворюють базу в~$\mathds E^4$.

}
%%%%%%%%%%%%%%%%%%%%%%%%%%%%%%%%%%%%%%%%%%%%%%%%%%%%%%%%%%%%%%%%%%%%%%

%       АНГЛІЙСЬКА ЧАСТИНА

\bigskip

\begin{center}
{\bf THIRD ORDER VARIATIONAL EQUATION FOR THE FREE RELATIVISTIC TOP}

\bigskip

{\it Roman~MATSYUK}

\medskip
Institute for Applied Problems in Mechanics and Mathematics\\3$^{\mbox b}$~Naukova~St., L\kern-.25em'\kern.04emviv, Ukraine

\end{center}

\medskip

I proffer a development of some third order equation of motion for the free relativistic top from the simultaneously imposed assumptions of variationality and Lorentz symmetry.

\end{document}